\documentclass[aps,prl,twocolumn,amsmath,amssymb,nofootinbib]{revtex4-2}
\usepackage{graphicx}
\usepackage{bm}
\usepackage{mathtools}
\usepackage{booktabs}
\usepackage{xcolor}
\usepackage{placeins}
\usepackage{hyperref}
\graphicspath{{../figures/}}
\hypersetup{colorlinks=true,linkcolor=blue!55!black,citecolor=green!40!black,urlcolor=blue!60!black}
\newcommand{\dd}{\mathrm d}
\newcommand{\tr}{\operatorname{tr}}
\newcommand{\qchi}{q_{\chi}}
\newcommand{\Astar}{\mathsf A_{*}}
\newcommand{\yast}{y_{*}^{(\chi)}}
\allowdisplaybreaks[3]

\begin{document}

\title{Shear No-Go Theorem and Poiseuille Focusing 
of Chiral Active Particles}

\author{Yating Wang}
\affiliation{School of Physics and Electronic Information,
Inner Mongolia Normal University, Hohhot 010022, China}
\author{Hao Wu}
\email{wuhao@ucas.ac.cn}
\affiliation{Zhejiang Key Laboratory of Soft Matter Biomedical Materials, Wenzhou Institute, University of Chinese Academy of Sciences,
Wenzhou, Zhejiang 325000, China}

\date{\today}

\begin{abstract}
Chiral active particles are widely expected to separate laterally in shear flow, but the symmetry requirements for such a current remain elusive. We prove a strict no-go theorem for uniform linear shear: the angular speed is \(\pi\) periodic while the transverse swimming velocity is antiperiodic, forbidding any net transverse displacement over a complete orientational cycle. Constant intrinsic spin and symmetric noise preserve this cancellation. We show that pressure-driven Poiseuille flow breaks this symmetry and generates a non-conservative focusing mechanism for mirror enantiomers. Crucially, this system exhibits a self-similar structure: despite the multi-parameter nature of the geometry and flow, the microscopic details undergo parameter decoupling and collapse into a single scaling combination. This yields a universal scaling law for the fixed-point position, \(y_{*}^{(\chi)} \sim \omega^{1/(n-1)}\), defining a universality class governed solely by the local curvature of the flow profile. A striking data collapse unifies various flow geometries and rates. An Ornstein--Uhlenbeck analysis provides the noise-limited bandwidth, while three-dimensional hydrodynamic calculations confirm the required even/odd chiral parity. These results establish predictive scaling laws and correct bounded-channel observables for chiral sorting.
\end{abstract}

\maketitle


At low Reynolds number, the translational and rotational velocities of a rigid particle follow instantaneously from the Stokes boundary-value problem. For an anisotropic body, the background strain couples to rotation and produces Jeffery-type dynamics rather than co-rotation with the local vorticity \cite{Jeffery1922,HappelBrenner,KimKarrila}. Active propulsion, intrinsic spin, thermal fluctuations, and confinement generate a broad set of non-equilibrium trajectories \cite{tenHagen2011,Wittkowski2012,ZottlStark2012,KhatriBurada2022,Valani2024}. Chirality introduces an additional layer: it permits pseudoscalar response coefficients that are forbidden in achiral systems. In chiral active matter, this symmetry breaking manifests as pseudoscalar transport coefficients and chiral transverse responses that are absent in achiral systems \cite{Mecke2024,Markovich2024}. A natural question is whether such pseudoscalar responses can be harnessed to produce a transverse current in flow.

For passive chiral objects, this question has a well-established answer. Fu et~al. showed that a helical bacterium in Poiseuille flow drifts perpendicular to the shear plane, providing a hydrodynamic route to chiral separation \cite{Fu2009}. For active chiral particles, the situation is more subtle. Experiments and simulations have demonstrated chiral-dependent orientation dynamics and biased locomotion in shear flow \cite{Jing2020,Zottl2023,Ogawa2025}, and these observations are often discussed as a basis for microfluidic chiral sorting. Recent reviews have systematized the microscopic separation mechanisms and their translation into macroscopic visualization \cite{Niu2025}. However, translating these mechanisms into a predictive framework faces a fundamental challenge: the existence of chirality-dependent response coefficients is not by itself sufficient to produce a transverse current. The candidate current must also be allowed by the exact symmetries of the closed dynamical equations.

This work compares two flows to identify the symmetry requirements. In uniform linear shear the local shear rate is constant and the transverse coordinate does not enter the angular velocity. For a polar swimmer whose transverse velocity reverses under a half turn, a $\pi$-periodic angular speed gives an exact cycle-by-cycle cancellation. A constant active spin changes the locking boundaries but not this cancellation. Noise changes individual trajectories but not the path-pairing symmetry of an ensemble with symmetric initial data. Finite-time oscillations and arbitrary translation-rotation resistance blocks therefore do not establish persistent cross-stream sorting.

\begin{figure*}[t]
 \centering
 \includegraphics[width=\linewidth]{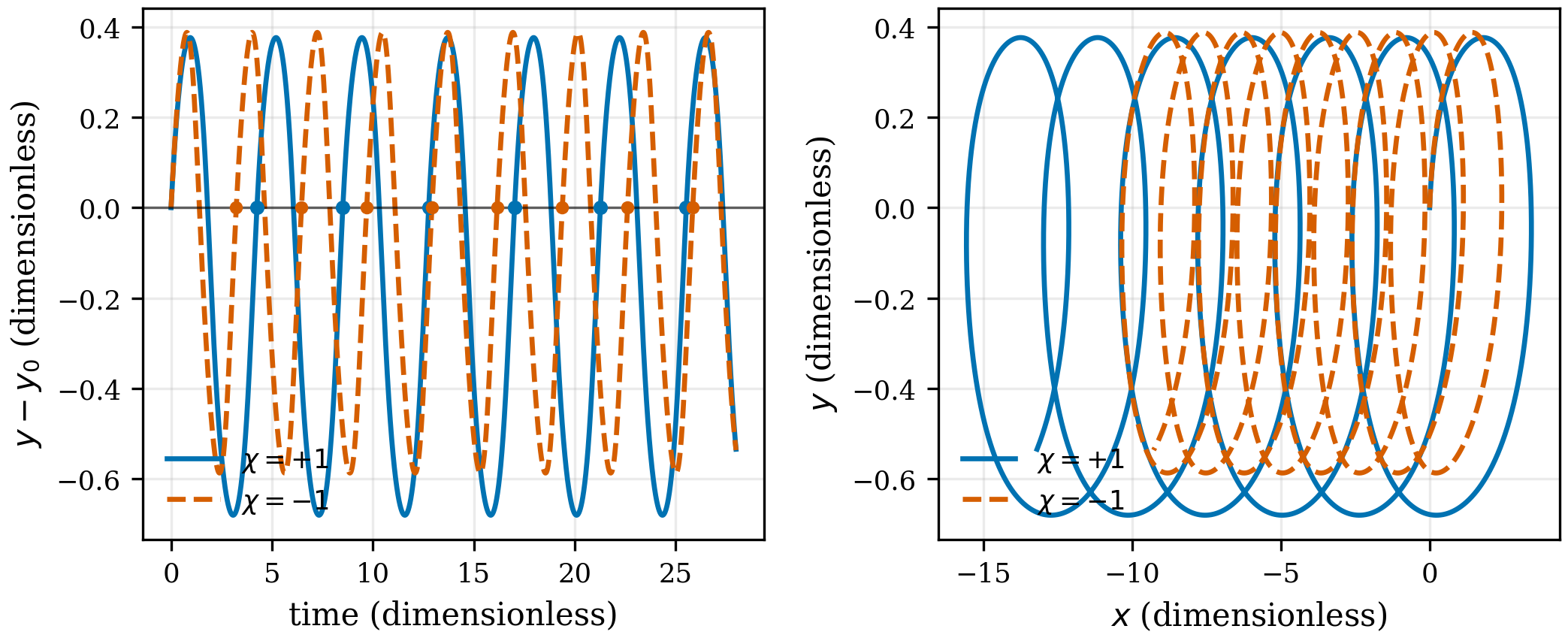}
 \caption{Uniform-shear trajectories oscillate in the transverse direction. The position returns after every complete polar cycle while the particle advances downstream. Both chiralities and both signs of intrinsic spin are shown.}
 \label{fig:uniform-cancel}
\end{figure*}

In pressure-driven Poiseuille flow, the local shear rate depends on the transverse position \cite{tenHagen2011,Asheichyk2019,PengBrady2020,GuptaBurada2025}, breaking the one-dimensional cancellation. The spinless position-orientation dynamics becomes integrable, supporting only neutral swinging orbits. We show that an attracting focus requires an additional non-conservative ingredient: intrinsic chiral spin combined with the anisotropic shape response phase. Crucially, the resulting off-center fixed points are not arbitrary; their positions obey a universal scaling law, $|y_*| \sim |\omega|^{1/(n-1)}$, dictated solely by the local curvature order $n$ of the flow profile. For standard parabolic flow ($n=2$), this yields an exact inverse-flow scaling, providing a robust, tunable observable for chiral sorting.

By reducing the full Stokes problem to a minimal hydrodynamic model, we establish a rigorous symmetry criterion for transverse drift and uncover a universality class for shear-driven chiral focusing. We derive the no-go theorem for uniform shear, the integrability of the spinless Poiseuille model, and the analytic scaling laws for the focusing position, time, and noise-limited bandwidth. Three-dimensional multiblob calibrations confirm the required even/odd parity of the chiral response coefficients. These results provide a predictive theoretical framework for designing chiral microfluidic sorting devices.

\paragraph{Minimal hydrodynamic model.} For a rigid particle in a background linear flow
$\bm u^{\infty}(\bm x)=\bm U^{\infty}+\bm\Omega^{\infty}\times\bm r
+\bm E^{\infty}\bm r$, the hydrodynamic force, torque, and stresslet relate to the rigid velocity and strain via a grand resistance matrix $\mathsf R$ \cite{HappelBrenner,KimKarrila}. A rigid-multiblob discretization captures the rigid kinematics \cite{BradyBossis1988,SierouBrady2001,ElfringBrady2022}, while a diagonal bead drag yields the free-draining limit (see End Matter~\hyperref[app:rpy]{A} for the saddle-point system and Rotne--Prager--Yamakawa (RPY) mobility \cite{RotnePrager1969,Yamakawa1970}).

The two-dimensional angular response to uniform shear is written in the general second-harmonic form
\begin{equation}
 \dot\theta=\chi\Omega_{0}
 -\frac{\Gamma}{2}\bigl[1-B\cos2\theta-\chi C\sin2\theta\bigr],
 \label{eq:angular-general}
\end{equation}
where $\chi=\pm1$ labels mirror chirality and $\Gamma$ is the local shear rate. This form follows from the reflection symmetry of the strain-coupling blocks of $\mathsf R$. Writing $B=\Lambda\cos\delta$ and $C=\Lambda\sin\delta$ yields $q_{\chi}(\theta) = 1-\Lambda\cos(2\theta-\chi\delta)$. Only the amplitude $\Lambda$ and the phase relative to the propulsion axis are invariant under redefinitions of the body zero angle. The even coefficient $B$ is mirror invariant, whereas the pseudoscalar phase coefficient $C$ changes sign under mirror reflection.

As a minimal analytical example, we consider three independent point-like beads. Exact torque balance yields \(B=-3/5\) and \(C=4\sqrt{3}/15\), with the translation-rotation coupling vanishing precisely at the friction center (see End Matter~\hyperref[app:rpy]{A}). In the following illustrations, we adopt \(B=0.45\) and \(C=-0.25\) to display the dynamical regimes clearly.

\paragraph{Shear no-go theorem.}\label{sec:nogo} After scaling the swimming speed to unity, the minimal uniform-shear model is
\begin{align}
 \dot x&=\Gamma y+\cos(\theta+\beta),\label{eq:usx}\\
 \dot y&=\sin(\theta+\beta),\label{eq:usy}\\
 \dot\theta&=\chi\omega-\frac{\Gamma}{2}\qchi(\theta),
 \label{eq:ustheta}
\end{align}
where $\omega \equiv \Omega_0$ is the intrinsic spin. The essential symmetry is
\begin{equation}
 \dot\theta(\theta+\pi)=\dot\theta(\theta),
 \quad
 \dot y(\theta+\pi)=-\dot y(\theta).
 \label{eq:halfsym}
\end{equation}
For an angular speed $f(\theta)$ that is continuous, nonzero, and $\pi$ periodic, and a transverse velocity $g(\theta)$ that is antiperiodic, the transverse displacement over a full polar cycle is
\begin{align}
 \Delta y
 =\int_{\theta_{0}}^{\theta_{0}+\pi}\frac{g(\theta)}{f(\theta)}\,\dd\theta
 +\int_{\theta_{0}}^{\theta_{0}+\pi}
 \frac{g(\vartheta+\pi)}{f(\vartheta+\pi)}\,\dd\vartheta = 0,
 \label{eq:nodrift-proof}
\end{align}
since $g(\vartheta+\pi)=-g(\vartheta)$ and $f(\vartheta+\pi)=f(\vartheta)$. 

Thus, every tumbling trajectory of Eqs.~\eqref{eq:usy}--\eqref{eq:ustheta} has zero cycle-averaged transverse velocity, regardless of chirality or constant intrinsic spin. For $\omega=0$, the phase-shifted Jeffery equation has the analytic half-turn period
\begin{equation}
 T_{\pi}=\frac{2\pi}{|\Gamma|\sqrt{1-\Lambda^{2}}},
 \label{eq:period}
\end{equation}
which we verify numerically in Fig.~\ref{fig:uniform-cancel}.

\begin{figure}[t]
 \centering
 \includegraphics[width=\linewidth]{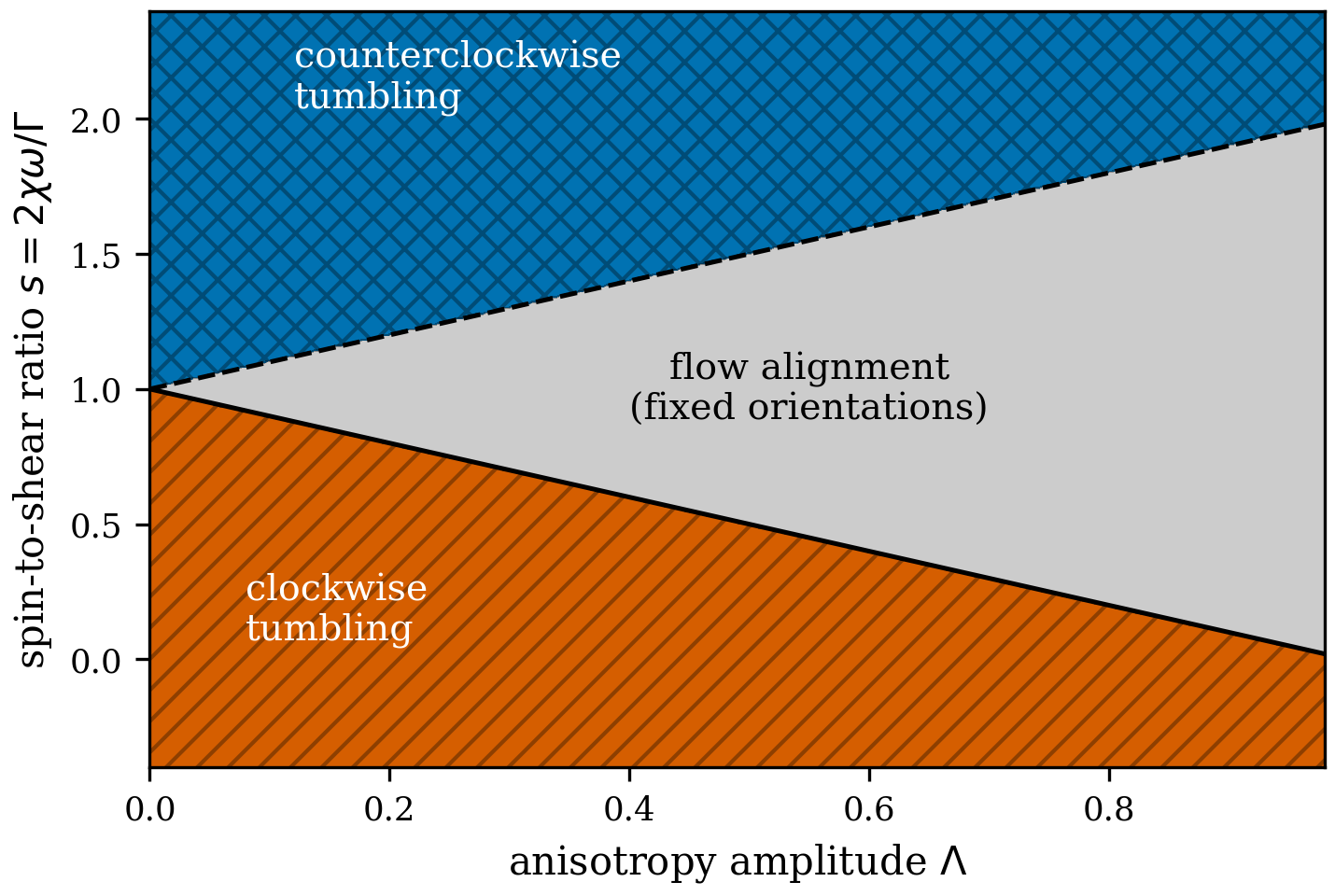}
 \caption{Uniform-shear rotational states. Constant intrinsic spin creates a locking wedge bounded by Eq.~\eqref{eq:locking}. Outside the wedge the cycle-cancellation result applies.}
 \label{fig:lockingphase}
\end{figure}

Defining $s = 2\chi\omega/\Gamma$, fixed orientations exist when
\begin{equation}
 1-\Lambda\le s\le1+\Lambda.
 \label{eq:locking}
\end{equation}
Outside this interval, the particle tumbles and Eq.~\eqref{eq:nodrift-proof} applies. Inside, an unbounded-domain model may have a constant transverse swimming component when the locked angle is not aligned with the flow; in a bounded channel, this state terminates at a wall (Fig.~\ref{fig:lockingphase}).

For rotational diffusion $\dd\theta=f_{\chi}(\theta)\dd t+\sqrt{2D_{r}}\,\dd W_{\theta}$, the Fokker--Planck (FP) generator is invariant under $(y,\theta)\mapsto(-y,\theta+\pi)$. A symmetric initial density remains symmetric, and all odd transverse moments vanish, even with an Ornstein--Uhlenbeck (OU) colored torque. An asymmetric initial orientation can generate a transient displacement, but the finite-time velocity decays as $O(T^{-1})$. The longitudinal displacement exhibits standard shear-enhanced scalings: passive Brownian motion gives $\langle\Delta x^{2}\rangle=2Dt+\tfrac{2}{3}D\Gamma^{2}t^{3}$, while a locked deterministic swimmer gives $\Delta x^{2}\sim t^{4}$. These longitudinal laws do not by themselves establish transverse sorting.

\paragraph{Poiseuille Focusing.} The no-go theorem relies on a constant local shear rate; in Poiseuille flow, this rate depends on the transverse position, circumventing the strict symmetry. Below, we show that the spinless model remains integrable, and a non-conservative spinning term is required to generate an attracting focus.

For a channel of half-width \(H\) and centerline speed \(U_m\), \(u_x^\infty(Y) = U_m(1-Y^2/H^2)\). Scaling length by \(H\) and time by \(H/v_s\) gives
\begin{align}
 \dot x&=\Pi(1-y^{2})+\cos(\theta+\beta),\label{eq:px}\\
 \dot y&=\sin(\theta+\beta),\label{eq:py}\\
 \dot\theta&=\chi\omega+\Pi y\qchi(\theta),
 \label{eq:ptheta}
\end{align}
where \(\Pi=U_m/v_s\) and \(\omega=\Omega_0 H/v_s\).

\begin{figure}[t]
 \centering
 \includegraphics[width=\linewidth]{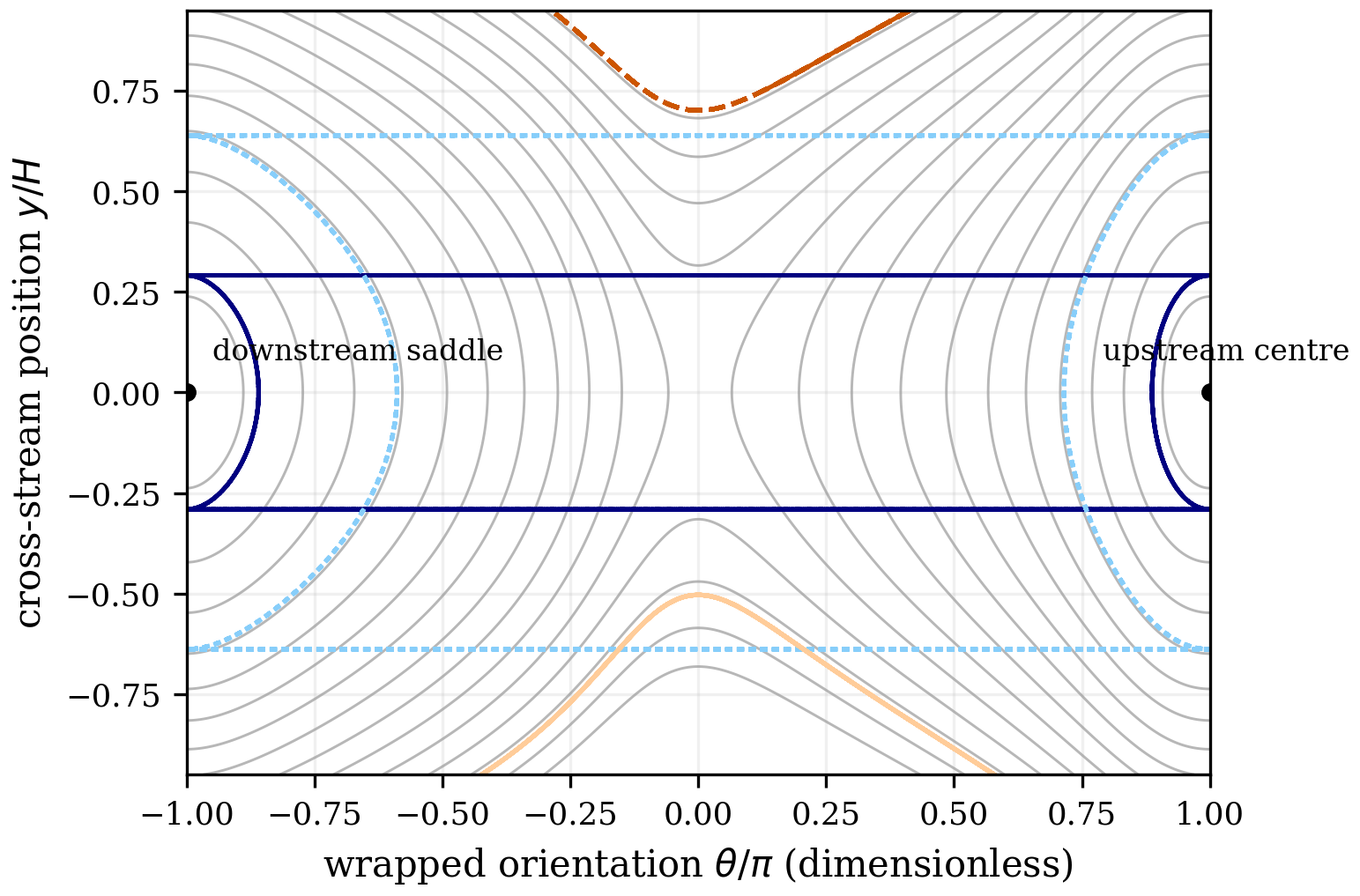}
 \caption{Spinless Poiseuille dynamics: conserved level sets, neutral upstream swinging, and a downstream saddle.}
 \label{fig:spinless}
\end{figure}

When \(\omega=0\), the flow admits the first integral
\begin{equation}
 \mathcal H_{\chi}(y,\theta)=\frac{y^{2}}{2}-\frac{I_{\chi}(\theta)}{\Pi},
 \
 I_{\chi}(\theta)=\int^{\theta}\frac{\sin(s+\beta)}{\qchi(s)}\,\dd s.
 \label{eq:H}
\end{equation}
Along the flow, \(\dd\mathcal H_{\chi}/\dd t=0\) (End Matter~\hyperref[app:integrability]{B}). The upstream state is a nonlinear center surrounded by neutral swinging orbits, and the downstream state is a saddle; there is no open basin of attraction. Fig.~\ref{fig:spinless} overlays numerical trajectories on the analytic level sets. For any steady parallel flow \(u(y)\bm e_x\), this construction yields the conserved quantity \(u(y)+2I_{\chi}(\theta)\). Thus, a spatially varying shear alone does not produce a bulk-focusing mechanism; a non-conservative spinning term is the second required ingredient.

\begin{figure*}[t]
 \centering
 \includegraphics[width=\linewidth]{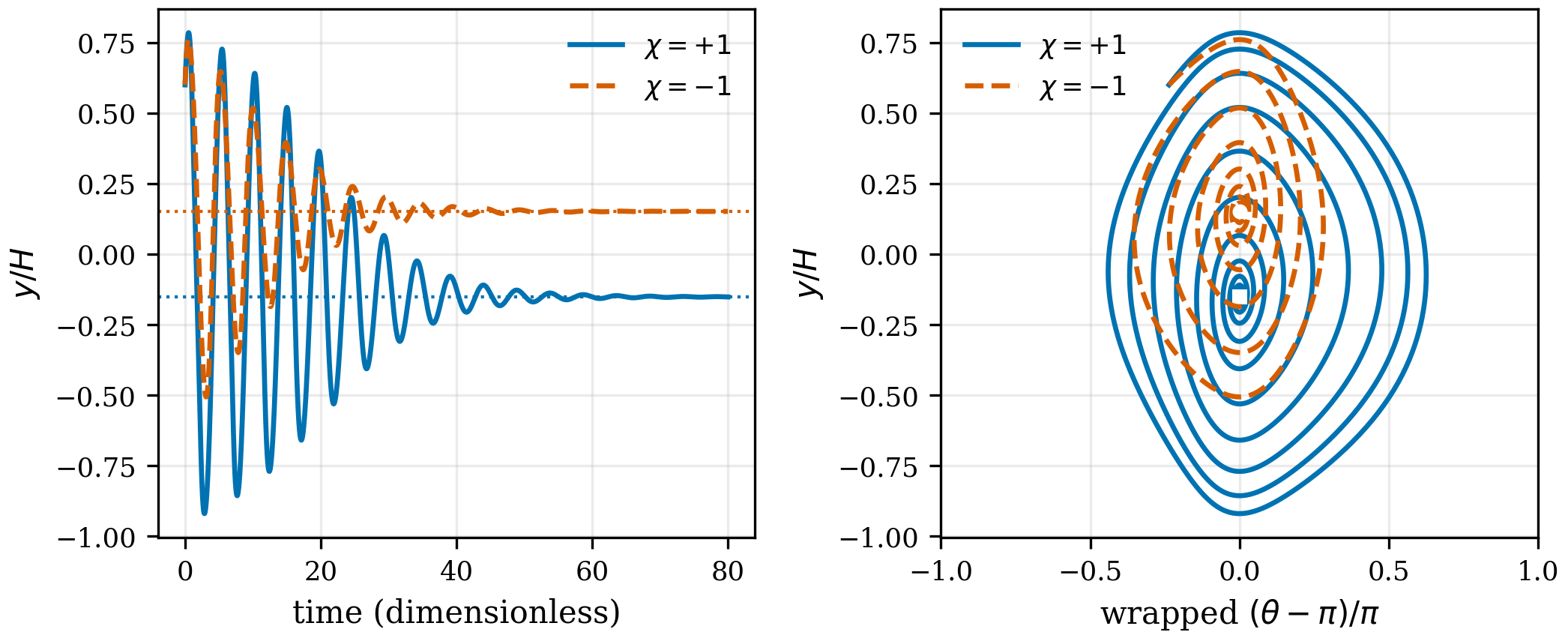}
 \caption{Poiseuille fixed points for the two sign conventions. Upper panel: an intrinsic mirror pair has \(\Omega_{\chi}=\chi\omega\) and \(C_{\chi}=\chi C_{0}\), so both states are stable and their positions are opposite. Lower panel: a same-sign applied torque leaves one enantiomer stable and does not produce opposite formal fixed points.}
 \label{fig:focusing}
\end{figure*}

Let the physical spin of chirality \(\chi\) be \(\Omega_{\chi}\). With \(C_{\chi}=\chi C_{0}\) and \(\beta=0\), the planar Poiseuille equations are
\begin{align}
 \dot y&=\sin\theta,\\
 \dot\theta&=\Omega_{\chi}+\Pi y \bigl[1-B\cos2\theta-\chi C_{0}\sin2\theta\bigr].
 \label{eq:paritymodel}
\end{align}
Fixed points require \(\sin\theta_*=0\), yielding the upstream branch \(\theta_*=\pi\):
\begin{equation}
 \yast=-\frac{\Omega_{\chi}}{\Pi(1-B)}.
 \label{eq:ystar-general-parity}
\end{equation}
Linearizing in \((y,\theta-\pi)\) yields the Jacobian \(\Astar\), with
\begin{equation}
 \tr\Astar=\frac{2\chi C_{0}\Omega_{\chi}}{1-B},
 \quad \det\Astar=\Pi(1-B).
 \label{eq:trace-general-parity}
\end{equation}
The eigenvalues follow from \(\lambda^{2}-(\tr\Astar)\lambda+\det\Astar=0\):
\begin{equation}
 \lambda_{\pm}=\frac{\chi C_{0}\Omega_{\chi}}{1-B} \pm\sqrt{\left(\frac{\chi C_{0}\Omega_{\chi}}{1-B}\right)^{2}-\Pi(1-B)}.
 \label{eq:eigs-general-parity}
\end{equation}
End Matter~\hyperref[app:linearization]{C} provides the Taylor expansion.

Two sign conventions are physically distinct. For an intrinsic mirror pair (\(\Omega_{\chi}=\chi\omega\), \(C_{\chi}=\chi C_0\)), the two enantiomers focus on opposite sides with identical relaxation rates:
\begin{equation}
 \yast=-\frac{\chi\omega}{\Pi(1-B)}, \quad \tr\Astar=\frac{2\omega C_{0}}{1-B}.
 \label{eq:intrinsic-parity-result}
\end{equation}
Both states are stable when \(\omega C_{0}<0\). Conversely, for a same-sign applied torque \(\Omega_{\chi}=\omega_{\mathrm{ext}}\), both formal fixed points share the same \(y_*\) but possess opposite traces, allowing only one enantiomer to be stable when \(C_0\omega_{\mathrm{ext}}\ne0\). This selective-capture protocol is distinct from the mirror-intrinsic two-band mechanism (Fig.~\ref{fig:focusing}).

\begin{figure}[t]
 \centering
 \includegraphics[width=\linewidth]{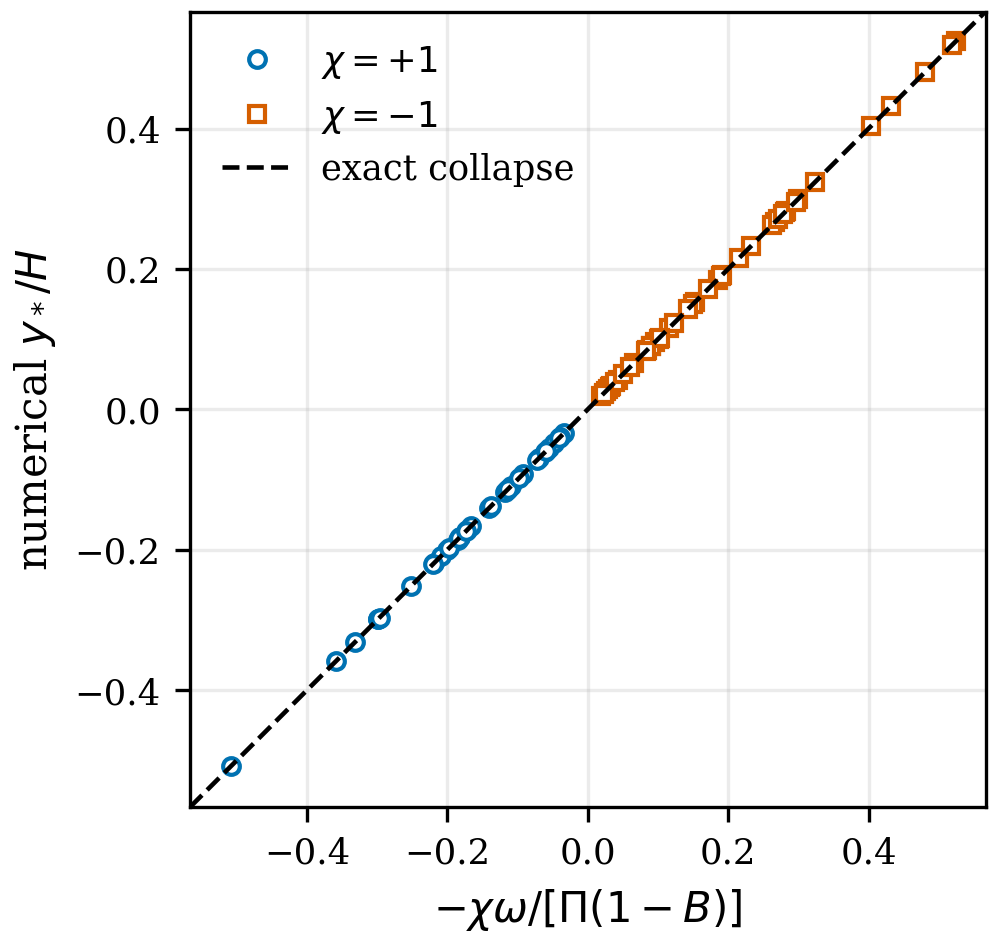}
 \caption{Fixed-point collapse for 80 parameter sets and both chiralities. The solid line is the analytic law \(y_{*}=-\chi\omega/[\Pi(1-B)]\).}
 \label{fig:collapse}
\end{figure}

Restoring dimensions in Eq.~\eqref{eq:intrinsic-parity-result} yields the peak separation and focusing time:
\begin{equation}
 \frac{Y_{*}^{(\chi)}}{H} =-\frac{\chi\Omega_{0}H}{U_{m}(1-B)}, \quad \Delta Y_{*} = \frac{2|\Omega_{0}|H^{2}}{U_{m}(1-B)},
 \label{eq:dimposition}
\end{equation}
\begin{equation}
 \tau_{f}=\frac{1-B}{|\Omega_{0}C|}.
 \label{eq:focustime}
\end{equation}
The weakly damped oscillation frequency scales as \(\Omega_{f,\mathrm{dim}}\sim H^{-1}\sqrt{U_{m}v_{s}(1-B)}\). The fixed point lies in the effective bulk region (\(|y|<y_w\)) provided \(|\omega|<\Pi(1-B)y_w\); outside this region, wall and finite-size effects dominate, which can be systematically quantified via rigid-multiblob methods \cite{Usabiaga2016}. Equation~\eqref{eq:intrinsic-parity-result} gives the parameter-free scaling \(y_{*}=-\chi\omega/[\Pi(1-B)]\). Eighty independent nonlinear root solves collapse onto this line with an rms error \(1.2\times10^{-15}\) (Fig.~\ref{fig:collapse}).

For a general steady parallel flow \(u(y)\bm e_x\), the spinless invariant holds. With intrinsic spin, the upstream fixed point satisfies
\begin{equation}
 u'(y_{*})=\frac{2\chi\omega}{q_{*}},
 \label{eq:generalprofile}
\end{equation}
where \(q_{*}=1-B\). For a centerline profile \(u(y)=u_0-A|y|^n\), we have \(u'(y)=-nA|y|^{n-1}\operatorname{sgn}(y)\). Substituting into Eq.~\eqref{eq:generalprofile} gives
\begin{equation}
 |y_{*}|^{n-1}=\frac{2|\omega|}{nA|q_{*}|} \implies |y_{*}|\sim|\omega|^{1/(n-1)}.
 \label{eq:general-scaling}
\end{equation}
Parabolic Poiseuille flow (\(n=2\)) gives \(|y_{*}|\propto|\omega|\), while a quartic profile (\(n=4\)) yields \(|y_{*}|\propto|\omega|^{1/3}\). This defines a universality class based on the local order of the velocity profile.

\paragraph{Noise-Limited Sorting.} With translational and rotational diffusion,
\begin{align}
 \dd y&=\sin\theta\,\dd t+\sqrt{2D_{y}}\,\dd W_{y},\\
 \dd\theta&=[\chi\omega+\Pi y\qchi(\theta)]\,\dd t+\sqrt{2D_{r}}\,\dd W_{\theta}.
 \label{eq:stochP}
\end{align}
The corresponding FP equation preserves the mirror symmetry \(p_{-\chi}^{\mathrm{ss}}(y,\theta)=p_{\chi}^{\mathrm{ss}}(-y,-\theta)\).

Near the stable upstream focus, let \(\eta=y-y_{*}\) and \(\phi=\theta-\pi\). Linearizing Eq.~\eqref{eq:stochP} gives the OU process,
\begin{equation}
 \dd\begin{pmatrix}\eta\\\phi\end{pmatrix} = \begin{pmatrix}0&-1\\ k&-\gamma\end{pmatrix}\begin{pmatrix}\eta\\\phi\end{pmatrix}\dd t + \begin{pmatrix}\sqrt{2D_{y}}&0\\0&\sqrt{2D_{r}}\end{pmatrix}\dd\bm W,
 \label{eq:OU}
\end{equation}
with \(k=\Pi(1-B)\) and \(\gamma=-2\omega C/(1-B)>0\). End Matter~\hyperref[app:linearization-Brownian]{D} provides the Taylor expansion leading to this linear system. The stationary covariance yields the transverse variance 
\begin{equation}
 \sigma_{y}^{2}=\frac{D_{y}(k+\gamma^{2})+D_{r}}{k\gamma}.
 \label{eq:cov}
\end{equation} 
as derived in End Matter~\hyperref[app:spectrum]{E} for the corresponding stationary power spectrum, which allows direct extraction of $k$ and $\gamma$ from measured trajectories. For \(D_{y}=0\), this reduces to \(\sigma_{y}^{2}=D_{r}/(-2\Pi\omega C)\). The resolution metric \(\mathcal S=|\mu_{+}-\mu_{-}|/\sqrt{\sigma_{+}^{2}+\sigma_{-}^{2}}\) becomes \(\mathcal S_{\mathrm{OU}}=\sqrt2|\omega|/(\Pi(1-B)\sigma_{y})\) in the mirror-symmetric OU approximation.

\begin{figure}[t]
 \centering
 \includegraphics[width=\linewidth]{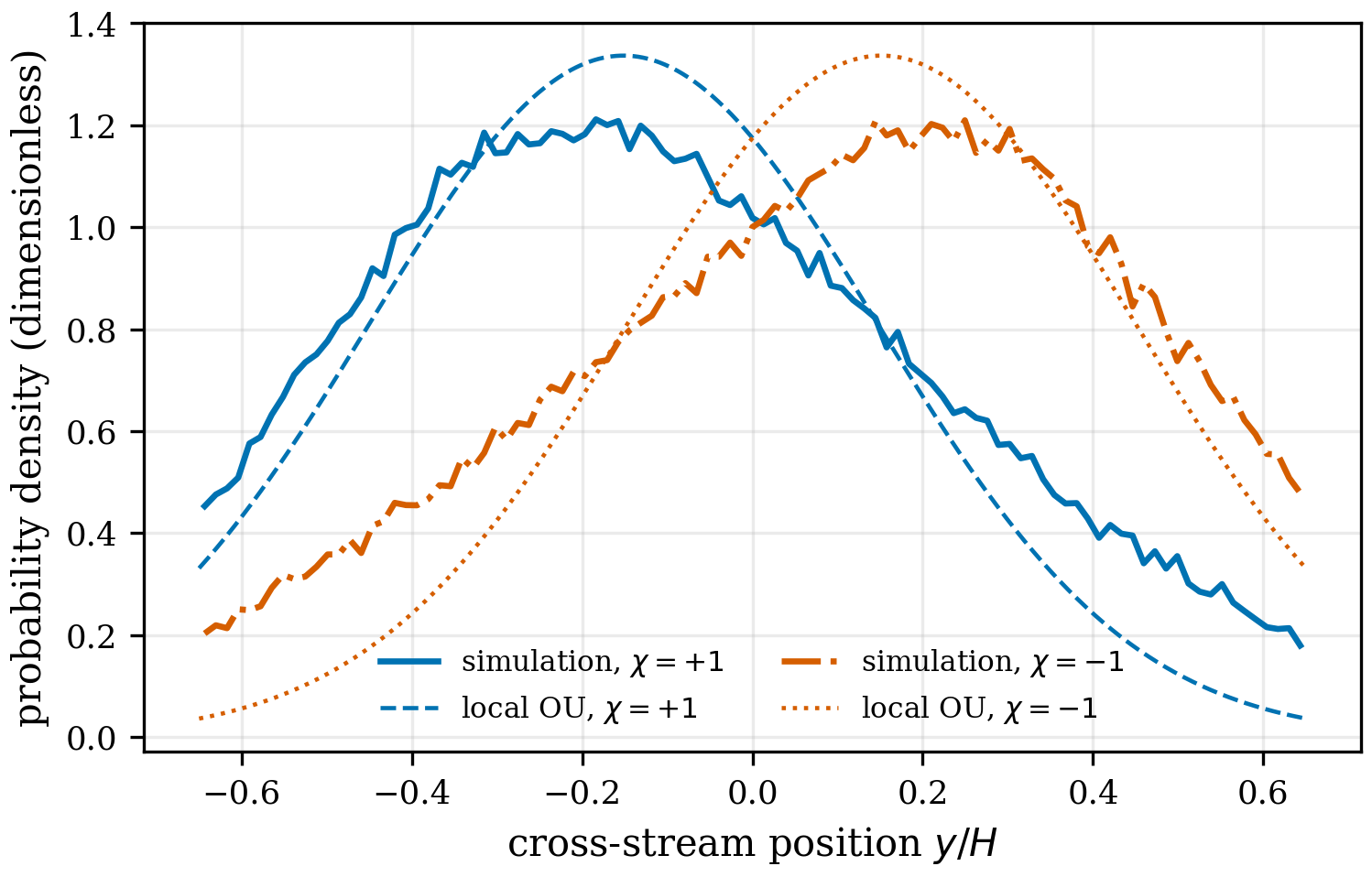}
 \caption{Brownian Poiseuille focusing. The nonlinear steady distributions are mirror related and overlap for the chosen noise levels. The deterministic centers are marked by vertical lines.}
 \label{fig:brownian}
\end{figure}

We perform nonlinear Brownian dynamics simulations using the Ermak–McCammon algorithm \cite{ErmakMcCammon1978}, which accounts for position-dependent mobility and thermal drift. The simulations reproduce the local Gaussian result at weak noise (Fig.~\ref{fig:brownian}). For representative parameters, the deterministic centers are \(\mp0.1515\), while simulated means are approximately \(-0.123\) and \(0.120\) with standard deviations \(0.367\) and \(0.373\). The strong overlap demonstrates noisy focusing rather than high-purity separation.

Every bounded trajectory satisfies \([y(T)-y(0)]/T=0\). Therefore, a sorting device must be characterized by its focused positions, bandwidths, establishment lengths, outlet enrichment, and throughput, rather than a nonzero infinite-time transverse velocity. 

To ground this parity structure in realistic 3D hydrodynamics, we perform a rigid-multiblob calculation \cite{RotnePrager1969,Yamakawa1970,Zuk2014} on an explicitly chiral particle: a twisted triangular frustum with unequal sides, alongside its exact mirror. The right-handed surface is generated by the reflection \(y\mapsto-y\). Volume, area, and principal moments agree to numerical roundoff, while an oriented landmark tetrahedron changes sign, confirming true geometric chirality (End Matter~\hyperref[app:rpy]{A}).

\paragraph{Discussion.} Our no-go theorem establishes a strict symmetry criterion for transverse drift in uniform shear: the half-turn transformation inherently pairs paths with opposite transverse currents. Constant active torque preserves this pairing without generating a tumbling-cycle drift, and symmetric noise preserves the ensemble symmetry.

In Poiseuille flow, the spatially varying shear breaks this pairing and permits an attracting focus. The focus requires both intrinsic chiral spin and a phase-shifted anisotropic shape response. Because the spinless model is integrable, a nonzero $C$ without spin leaves the linearization trace zero. The mirror-pair focusing is therefore a collective, non-conservative consequence of chirality, activity, and flow curvature.

Most importantly, we uncover a universality class governed by the local velocity profile order \(n\). This reveals a self-similar structure reminiscent of renormalization-group scale invariance: despite the multi-parameter nature of the system, the microscopic details decouple into a single scaling combination \(\Xi = \chi \Omega_0 H / [U_m(1-B)]\). As shown in Fig.~\ref{fig:collapse}, eighty independent nonlinear root solves undergo a data collapse onto the universal line \(y_* = \Xi\). This scaling collapse yields the universal law $|y_{*}|\sim|\omega|^{1/(n-1)}$, unifying various flow geometries: parabolic flow (\(n=2\)) gives \(|y_{*}|\propto|\omega|\), while a quartic profile ($n=4$) gives \(|y_{*}|\propto|\omega|^{1/3}\). Furthermore, the focusing time \(\tau_{f}=(1-B)/|\Omega_{0}C|\) and the pure-rotational Brownian bandwidth \(\sigma_{y}^{2}=D_{r}/(-2\Pi\omega C)\) provide analytic predictions independent of flow rate. 

Consequently, the correct bounded-channel observables are focused positions, bandwidths, establishment lengths, outlet enrichment, and throughput. Every bounded trajectory satisfies \(\lim_{T\to\infty}[y(T)-y(0)]/T=0\), indicating that a nonzero infinite-time transverse velocity is fundamentally inconsistent as a sorting metric. 

Finally, extending the present rigid-body framework to deformable chiral microswimmers is a natural next step. Studies of amoeboid swimming have shown that shape deformation and confinement can qualitatively alter swimming speed, stresslet type, and trajectory stability \cite{Wu2015,Wu2016}. Allowing the propulsion and resistance coefficients to vary over an internal deformation cycle would make it possible to examine how shape dynamics modify the cancellation and focusing conditions identified here.

\paragraph{Conclusions.} By reducing the full Stokes problem to a minimal hydrodynamic model, this work establishes two fundamental advances for chiral active matter. First, we prove a rigorous no-go theorem for uniform linear shear, demonstrating that symmetric noise and constant intrinsic torque cannot generate a persistent transverse current. Second, we identify a non-conservative focusing mechanism in Poiseuille flow and uncover a universality class dictated by the local flow curvature.

We derive exact, predictive scaling laws for the mirror-enantiomer peak separation and focusing time. Through the scaling combination of physical parameters, we demonstrate a self-similar structure across different flow profiles. This parameter decoupling produces a universal data collapse--independent of specific flow rates, geometries, or chiralities--in the fixed-point position, linking microscopic chirality to macroscopic transport. A positive-definite RPY multiblob calculation on an explicit mirror geometry independently confirms the required even/odd parity of the chiral coefficients. These findings provide a testable theoretical framework and design principles for chiral microfluidic sorting devices, demonstrating that the universal interplay of shape anisotropy, activity, and flow curvature governs the spatial organization of chiral particles.

\paragraph{Acknowledgments.} We thank E. Frey and H. Stark for helpful discussions. Y.W. is supported by the Natural Science Foundation of Inner Mongolia Autonomous Region under Grant No. 2026LHQC0206 and the Science and Technology Plan of the Inner Mongolia under Grant No. 2026KYPT0079. H.W. is supported by the General Program of the National Natural Science Foundation of China (NSFC) under Grant No. 12374210 and the Open Research Fund of Songshan Lake Materials Laboratory under Grant No. 2023SLABFN20.

\clearpage

\onecolumngrid

\vspace{12pt}
\noindent\hrulefill \hspace{24pt} {\bf End Matter} \hspace{24pt} \hrulefill
\vspace{12pt}

\twocolumngrid

\renewcommand{\thesection}{\Alph{section}}
\setcounter{section}{0}

\section{A. RPY multiblob details}
\label{app:rpy}

\renewcommand{\theequation}{A\arabic{equation}}
\setcounter{equation}{0}
\renewcommand{\thefigure}{A\arabic{figure}}
\setcounter{figure}{0}

\begin{figure*}[t]
 \centering
 \includegraphics[width=\linewidth]{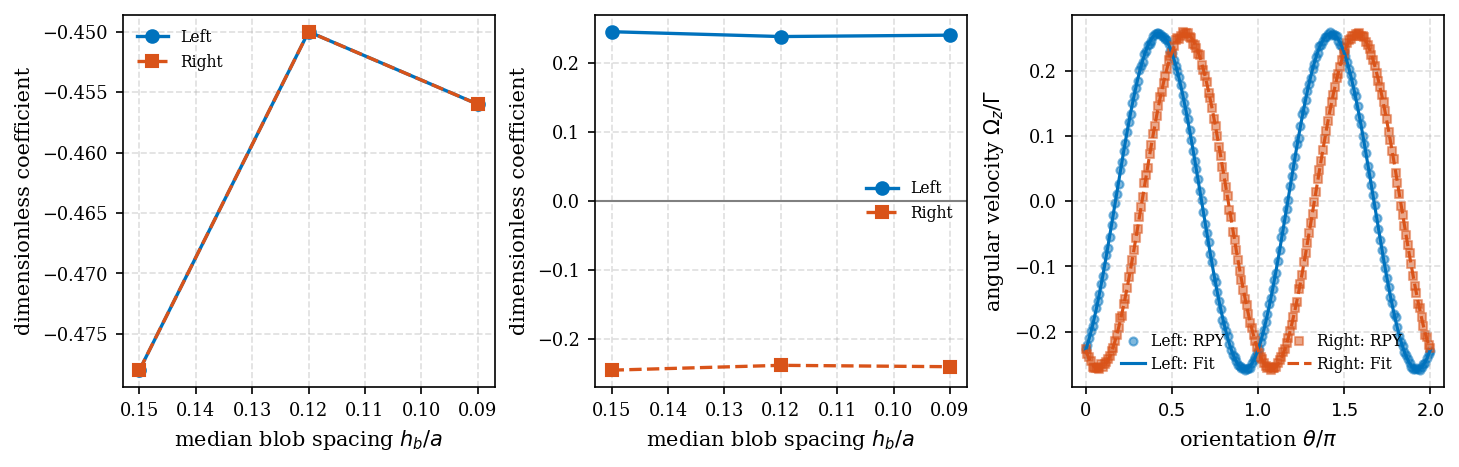}
 \caption{Three-resolution RPY extraction. The horizontal axis is the median blob spacing $h_b/a$ (decreasing from left to right, corresponding to increasing resolution). The even coefficient \(B\) is mirror invariant, the pseudoscalar phase coefficient \(C\) changes sign, and the rightmost panel demonstrates that the angular response is dominated by the second harmonic to numerical precision.}
 \label{fig:BC}
\end{figure*}

As a minimal analytical example, we consider three beads with the free-draining limit, where hydrodynamic interactions are neglected. The friction center of this geometry is located at
$\bm r_{c}=(-a/12,\sqrt3 a/6)$, and the weighted second moment about this point is
\begin{equation}
 \bm S=a^{2}
 \begin{pmatrix}
 3/16&-\sqrt3/8\\
 -\sqrt3/8&3/4
 \end{pmatrix}.
\end{equation}
By evaluating the torque balance in a shear flow, we obtain
\begin{equation}
 B=-\frac35,\quad C=\frac{4\sqrt3}{15},\quad
 \Lambda=\frac{\sqrt{129}}{15}.
 \label{eq:threebeadBC}
\end{equation}
This calculation confirms that the translation-rotation block vanishes exactly at the friction center, but remains nonzero if evaluated at the geometric origin.

The rigid-multiblob discretization represents the particle by blob forces $\bm\lambda$ and solves
\begin{equation}
 \begin{pmatrix}\bm M&-\bm K\\ \bm K^{\mathsf T}&0\end{pmatrix}
 \begin{pmatrix}\bm\lambda\\ \mathcal U\end{pmatrix}
 =
 \begin{pmatrix}-\bm u^{\infty}-\bm u^{\mathrm{slip}}\\
 \mathcal F^{\mathrm{ext}}\end{pmatrix},
 \label{eq:multiblob-app}
\end{equation}
where $\bm M$ contains all blob--blob hydrodynamic interactions and $\bm K$
imposes rigid kinematics. A diagonal collection of independent bead drags is a
free-draining approximation. The unbounded translational RPY blocks are
\begin{equation}
 \mathsf M_{ii}=\frac{\mathsf I}{6\pi\eta a_{b}},
\end{equation}

For equal blob radius $a_{b}$, the unbounded translational RPY blocks are
\begin{equation}
 \mathsf M_{ii}=\frac{\mathsf I}{6\pi\eta a_{b}},
\end{equation}
\begin{equation}
 \mathsf M_{ij}=\frac{1}{8\pi\eta r}
 \left[\left(1+\frac{2a_{b}^{2}}{3r^{2}}\right)\mathsf I
 +\left(1-\frac{2a_{b}^{2}}{r^{2}}\right)\hat{\bm r}\hat{\bm r}\right],
 \ r\ge2a_{b},
 \label{eq:rpy-far}
\end{equation}
with the regularized overlap branch
\begin{equation}
 \mathsf M_{ij}=\frac{1}{6\pi\eta a_{b}}
 \left[\left(1-\frac{9r}{32a_{b}}\right)\mathsf I
 +\frac{3r}{32a_{b}}\hat{\bm r}\hat{\bm r}\right],
 \ r<2a_{b}.
 \label{eq:rpy-near}
\end{equation}
The dense mobility is symmetric to roundoff and positive definite in the overlap
test. For a sphere represented by 322 blobs, the translational and rotational
resistance errors are $1.62\%$ and $5.30\%$, respectively.

For each surface resolution, the force- and torque-free angular velocity in unit
simple shear is fitted to
\begin{equation}
 \Omega_{z}(\theta)=-\frac12+\frac12\bigl[B\cos2\theta+C\sin2\theta\bigr].
\end{equation}
The constant coefficient equals $-1/2$ to $3.4\times10^{-16}$, and the fitted
fourth harmonic is below $2.2\times10^{-16}$. At the finest $N_{b}=220$
resolution,
\begin{equation}
 B_{L}=B_{R}=-0.455995,
 \
 C_{L}=-C_{R}=0.240371.
 \label{eq:stage2BC-app}
\end{equation}
The mirror parity errors are below $2\times10^{-16}$. The even coefficient $B$ is mirror invariant, the pseudoscalar phase coefficient $C$ changes sign, and the angular response is second harmonic to numerical precision
(Fig.~\ref{fig:BC}).

A tangential slip consisting of the surface projection of a body-fixed propulsion
direction plus a tangential swirl is prescribed. Reflecting both the mesh and
this polar slip field gives equal $U_{x}$ and speed, opposite $U_{y}$, opposite
$\Omega_{0}$, and opposite propulsion angle. For $N_{b}=220$ the result is
\begin{equation}
 (U_{x},U_{y},\Omega_{0})_{L}=(0.82465,-0.03813,-0.48837),
\end{equation}
with the mirror signs for the right-handed particle.

\section{B. Integrability of the spinless Poiseuille model}
\label{app:integrability}

\renewcommand{\theequation}{B\arabic{equation}}
\setcounter{equation}{0}
\renewcommand{\thefigure}{B\arabic{figure}}
\setcounter{figure}{0}

This section verifies that Eq.~\eqref{eq:H} is a first integral of the spinless
Poiseuille equations
\begin{align}
 \dot y&=\sin\theta,\\
 \dot\theta&=\Pi y\qchi(\theta),
\end{align}
obtained from Eqs.~\eqref{eq:py}--\eqref{eq:ptheta} with $\omega=0$ and
$\beta=0$. The first integral is
\begin{equation}
 \mathcal H_{\chi}(y,\theta)=\frac{y^{2}}{2}-\frac{I_{\chi}(\theta)}{\Pi},
 \quad
 I_{\chi}(\theta)=\int^{\theta}\frac{\sin s}{\qchi(s)}\,\dd s.
\end{equation}
Differentiating along the flow gives
\begin{align}
 \frac{\dd\mathcal H_{\chi}}{\dd t}
 =y\sin\theta
 -\frac{1}{\Pi}\frac{\sin\theta}{\qchi(\theta)}
 \bigl[\Pi y\qchi(\theta)\bigr].
 \label{eq:Hproof-app}
\end{align}
The two terms in Eq.~\eqref{eq:Hproof-app} are identical, so
$\dd\mathcal H_{\chi}/\dd t=0$.

For a general steady parallel flow $u(y)\bm e_{x}$ with
$\dot\theta=-u'(y)\qchi(\theta)/2$, the quantity
\begin{equation}
 \mathcal K=u(y)+2\int^{\theta}\frac{\sin s}{\qchi(s)}\,\dd s
\end{equation}
is conserved. Differentiating gives
\begin{align}
 \frac{\dd\mathcal K}{\dd t}
 =u'(y)\sin\theta
 +2\frac{\sin\theta}{\qchi(\theta)}
 \left[-\frac{u'(y)\qchi(\theta)}{2}\right]=0.
\end{align}
The conserved level sets of $\mathcal H_{\chi}$ and $\mathcal K$ define the
neutral swinging orbits used in the main text.

\section{C. Linearization of the deterministic spinning fixed point}
\label{app:linearization}

\renewcommand{\theequation}{C\arabic{equation}}
\setcounter{equation}{0}
\renewcommand{\thefigure}{C\arabic{figure}}
\setcounter{figure}{0}

This section derives the Jacobian and the eigenvalues of
Sec.~\hyperref[sec:nogo]{III} for the spinning Poiseuille model
\begin{align}
 \dot y&=\sin\theta,\\
 \dot\theta&=\Omega_{\chi}+\Pi y
 \bigl[1-B\cos2\theta-\chi C_{0}\sin2\theta\bigr].
 \label{eq:paritymodel-app}
\end{align}
At the upstream branch $\theta_{*}=\pi$, $\cos2\theta_{*}=1$ and
$\sin2\theta_{*}=0$, so
\begin{equation}
 0=\Omega_{\chi}+\Pi y_{*}(1-B),
\end{equation}
which gives Eq.~\eqref{eq:ystar-general-parity}.

Write $y=\yast+\eta$ and $\theta=\pi+\phi$. The trigonometric factors expand as
\begin{align}
 \sin(\pi+\phi)&=-\sin\phi=-\phi+O(\phi^{3}),\\
 \cos(2\pi+2\phi)&=\cos2\phi=1-2\phi^{2}+O(\phi^{4}),\\
 \sin(2\pi+2\phi)&=\sin2\phi=2\phi+O(\phi^{3}).
\end{align}
Substituting into Eq.~\eqref{eq:paritymodel-app} gives
\begin{align}
 \dot\eta=&-\phi+O(\phi^{3}),\\
 \dot\phi=&\Omega_{\chi}+\Pi(\yast+\eta)
 \bigl[1-B(1-2\phi^{2}+O(\phi^{4})) \\
 &-\chi C_{0}(2\phi+O(\phi^{3}))\bigr]. \nonumber
\end{align}
The zeroth-order term vanishes by the fixed-point condition. Retaining terms
linear in $\eta$ and $\phi$ gives
\begin{align}
 \dot\eta&=-\phi,\\
 \dot\phi&=\Pi(1-B)\eta
 +2\chi C_{0}\Pi\yast\phi.
\end{align}
Using Eq.~\eqref{eq:ystar-general-parity} for $\yast$,
\begin{equation}
 2\chi C_{0}\Pi\yast
 =-\frac{2\chi C_{0}\Omega_{\chi}}{1-B}.
\end{equation}
The linear system is therefore
\begin{equation}
\begin{aligned}
 \dd\begin{pmatrix}\eta\\\phi\end{pmatrix}
 &=
 \Astar
 \begin{pmatrix}\eta\\\phi\end{pmatrix}\dd t,
 \\
 \Astar&=
 \begin{pmatrix}
 0&-1\\
 \Pi(1-B)&2\chi C_{0}\Omega_{\chi}/(1-B)
 \end{pmatrix}.
\end{aligned}
\end{equation}
The trace and determinant are
\begin{equation}
 \tr\Astar=\frac{2\chi C_{0}\Omega_{\chi}}{1-B},
 \quad
 \det\Astar=\Pi(1-B).
\end{equation}
The characteristic polynomial
$\lambda^{2}-(\tr\Astar)\lambda+\det\Astar=0$ gives
\begin{equation}
 \lambda_{\pm}=\frac{\chi C_{0}\Omega_{\chi}}{1-B}
 \pm\sqrt{\left(\frac{\chi C_{0}\Omega_{\chi}}{1-B}\right)^{2}
 -\Pi(1-B)}.
\end{equation}
For an intrinsic mirror pair, $\Omega_{\chi}=\chi\omega$ and
$C_{\chi}=\chi C_{0}$, the trace becomes $2\omega C_{0}/(1-B)$ and the
eigenvalues reduce to 
\begin{equation}
 \lambda_{\pm}=\frac{\omega C_{0}}{1-B}
 \pm\sqrt{\left(\frac{\omega C_{0}}{1-B}\right)^{2}-\Pi(1-B)}.
 \label{eq:eigs}
\end{equation}
The stability condition $\tr\Astar<0$ gives $\omega C_{0}<0$. For a same-sign applied torque, $\Omega_{\chi}=\omega_{\mathrm{ext}}$, the trace is $2\chi C_{0}\omega_{\mathrm{ext}}/(1-B)$, so the two enantiomers have opposite traces and only one is stable.

\section{D. Linearization of the stochastic Poiseuille model}
\label{app:linearization-Brownian}

\renewcommand{\theequation}{D\arabic{equation}}
\setcounter{equation}{0}
\renewcommand{\thefigure}{D\arabic{figure}}
\setcounter{figure}{0}

This section expands Eq.~\eqref{eq:stochP} around the stable upstream focus.
Write $y=\yast+\eta$ and $\theta=\pi+\phi$. The drift terms are
\begin{align}
 &\sin(\pi+\phi)=-\phi+O(\phi^{3}),\\
 &\chi\omega+\Pi(\yast+\eta)\qchi(\pi+\phi) \notag\\
 =&\chi\omega+\Pi(\yast+\eta)
 \bigl[(1-B)-2\chi C_{0}\phi+O(\phi^{2})\bigr].
\end{align}
The zeroth-order term vanishes by the fixed-point condition
$\chi\omega+\Pi\yast(1-B)=0$. Retaining terms linear in $\eta$ and $\phi$ gives
\begin{align}
 \dot\eta&=-\phi,\\
 \dot\phi&=\Pi(1-B)\eta
 -2\Pi\yast\chi C_{0}\phi.
\end{align}
For the intrinsic mirror pair, $\yast=-\chi\omega/[\Pi(1-B)]$, so
\begin{equation}
 -2\Pi\yast\chi C_{0}
 =\frac{2\omega C_{0}}{1-B}.
\end{equation}
Since $\omega C_{0}<0$, define $\gamma=-2\omega C_{0}/(1-B) >0, k=\Pi(1-B)$. The linearized stochastic equations are therefore
\begin{align}
 \dd\eta&=-\phi\,\dd t+\sqrt{2D_{y}}\,\dd W_{y},\\
 \dd\phi&=(k\eta-\gamma\phi)\,\dd t+\sqrt{2D_{r}}\,\dd W_{\theta},
\end{align}
which is Eq.~\eqref{eq:OU}.

For the nonlinear Brownian simulations, the stochastic differential
equations are integrated with the EM algorithm
\cite{ErmakMcCammon1978}, a first-order Brownian dynamics scheme. In the
position-dependent-mobility form, the scheme includes the divergence of the mobility matrix, i.e., the It{\^o} drift term $k_BT\nabla\cdot\mathsf N$, which is required for correct equilibrium statistics. 

\section{E. OU position spectrum}
\label{app:spectrum}

\renewcommand{\theequation}{E\arabic{equation}}
\setcounter{equation}{0}
\renewcommand{\thefigure}{E\arabic{figure}}
\setcounter{figure}{0}

For the stable linearized system, the stationary transverse power spectrum is
\begin{equation}
 S_{y}(\nu)=
 \frac{2D_{r}+2D_{y}(\gamma^{2}+\nu^{2})}
 {(k-\nu^{2})^{2}+\gamma^{2}\nu^{2}}.
\end{equation}
This Lorentzian-like spectrum exhibits a damped-swinging peak at $\nu \approx \sqrt{k-\gamma^2/2}$. Measuring this peak directly from a single-particle trajectory provides a robust experimental route to infer $k$ and $\gamma$ without fitting the time-domain envelope.

\end{document}